\documentclass[10pt,aps,prb,twocolumn,superscriptaddress,showpacs,showkeys]{revtex4-1} 
\usepackage{graphicx}
\usepackage{graphicx,epstopdf,color}
\usepackage{amsfonts}
\usepackage{amsmath,amssymb,mathrsfs}
\usepackage{bm}
\usepackage{pst-grad}
\begin{document}

\title{Thermodynamic Anomaly Above the Superconducting Critical Temperature in the Quasi One-Dimensional Superconductor Ta$_4$Pd$_3$Te$_{16}$}

\author{T. Helm}
\affiliation{Department of Physics, University of California, Berkeley, California 94720, USA}
\affiliation{Materials Science Division, Lawrence Berkeley National Laboratory, Berkeley, California 94720, USA}
\author{F. Flicker}
\affiliation{Department of Physics, University of California, Berkeley, California 94720, USA}
\author{R. Kealhofer}
\affiliation{Department of Physics, University of California, Berkeley, California 94720, USA}
\author{P. Moll}
\affiliation{Department of Physics, University of California, Berkeley, California 94720, USA}
\affiliation{Materials Science Division, Lawrence Berkeley National Laboratory, Berkeley, California 94720, USA}
\author{I. M. Hayes}
\affiliation{Department of Physics, University of California, Berkeley, California 94720, USA}
\affiliation{Materials Science Division, Lawrence Berkeley National Laboratory, Berkeley, California 94720, USA}
\author{N. P. Breznay}
\affiliation{Department of Physics, University of California, Berkeley, California 94720, USA}
\affiliation{Materials Science Division, Lawrence Berkeley National Laboratory, Berkeley, California 94720, USA}
\author{Z. Li}
\affiliation{Department of Physics, University of California, Berkeley, California 94720, USA}
\author{S. G. Louie}
\affiliation{Department of Physics, University of California, Berkeley, California 94720, USA}
\author{Q. R. Zhang}
\affiliation{ National High Magnetic Field Laboratory, Florida State University, Tallahassee, Florida 32310, USA}
\author{L. Balicas}
\affiliation{ National High Magnetic Field Laboratory, Florida State University, Tallahassee, Florida 32310, USA}
\author{J. E. Moore}
\affiliation{Department of Physics, University of California, Berkeley, California 94720, USA}
\affiliation{Materials Science Division, Lawrence Berkeley National Laboratory, Berkeley, California 94720, USA}
\author{J. G. Analytis}
\affiliation{Department of Physics, University of California, Berkeley, California 94720, USA}
\affiliation{Materials Science Division, Lawrence Berkeley National Laboratory, Berkeley, California 94720, USA}

\pacs{74.70.-b,74.25.Jb, 71.18.+y, 74.25.Bt}

\begin{abstract}
We study the intrinsic transport anisotropy and fermiology of the quasi one-dimensional superconductor Ta$_4$Pd$_3$Te$_{16}$. Below $T^*=20$\,K we detect a thermodynamic phase transition that predominantly affects the conductivity perpendicular to the quasi one-dimensional chains, consistent with a thermodynamic transition related to the presence of charge order that precedes superconductivity. Remarkably the Fermi surface pockets detected by de Haas-van Alphen (dHvA) oscillations are unaffected by this transition, suggesting that the ordered state does not break any translational symmetries but rather alters the scattering of the quasiparticles themselves.
\end{abstract}

\keywords{Magnetic Quantum Oscillations, SdH and dHvA effect, Anisotropy, FIB, steady and pulsed magnetic fields, quasi one-dimensional superconductor, bulk Fermi surface,Charge density wave, CDW, phase transition, fluctuations}
\maketitle
\section{Introduction}

Superconducting materials with extreme anisotropies in their electronic normal state behavior are of immense interest for studying the relationship between dimensionality, broken symmetry, and unconventional mechanisms of superconductivity. For example, both cuprate and iron-pnictide superconductors are proximal to magnetic and possibly ``nematic" orders~\cite{Hinkov_2008,Fernandes_2014}. In the case of the cuprates, the recent discovery of charge density order has ignited debate as to whether such thermodynamic phases are competing, causal, or benignly coincident with superconductivity~\cite{Wu2011,Christensen2016,badoux2016}. This ground-state electronic competition is particularly important in quasi-one dimensional (q1D) systems~\cite{Peierls1930,Pouget2016}, which often show charge order and also exhibit unconventional superconductivity. Recently, the q1D ternary superconductor Ta$_4$Pd$_3$Te$_{16}$ (TPT) has been suggested to be proximal to quantum ordered states and a possible host for unconventional superconductivity~\cite{pan_nodal_2014, jiao_superconductivity_2014,du_anisotropic_2015}. 

TPT is a layered material with a monoclinic crystal structure~\cite{mar_synthesis_1991}. The high-conductivity direction coincides with the $b$ axis, parallel to the PdTe$_2$-chains lying within flat layers separated by TaTe$_3$ chains and Ta$_2$Te$_4$ double chains (see Fig~\ref{fig:FIB_Crys}a). Well-studied q1D systems like the organic TMTSF-PF$_6$~\cite{Dressel2007} or purple bronze Li$_{0.9}$Mo$_{6}$O$_{17}$~\cite{Wakeham2011} possess complex phase diagrams with evidences of a spin or charge density wave instability~\cite{Pouget2016}. However, despite several recent studies, TPT has not shown unambiguous evidence of an instability competing with its SC phase. This may be because TPT is significantly more isotropic: although direct resistivity measurements have been unavailable until this study, measurements of the upper critical field have given indirect evidence that the electronic anisotropy in this compound is relatively weak~\cite{pan_nodal_2014, jiao_superconductivity_2014}. Recent density functional theory (DFT) calculations predict that the system is multi-band in nature~\cite{singh_multiband_2014}, showing a combination of q2D, 3D and several q1D bands at the Fermi level $E_F$ that could account for earlier evidence of unconventional behavior in the thermal conductivity and heat capacity~\cite{pan_nodal_2014, jiao_superconductivity_2014}. Aside from superconductivity, no other bulk properties show clear thermodynamic evidence for an additional order parameter, raising the possibility that there may not be any other broken symmetry proximal to superconductivity. It is therefore important to determine the underlying Fermi surface as well as the intrinsic electronic anisotropy in order to understand the apparent absence of other ordered states in the case of TPT.

In this work we report the existence of a thermodynamic transition above the superconducting critical temperature ($T_c$), indicative of a nearby phase instability in TPT. By employing torque magnetometry and magnetotransport, we observed de Haas-van Alphen (dHvA) oscillations that confirm the multi-band nature of TPT's bulk Fermi surface. Using focused ion beam (FIB) microstructuring, we fabricated devices enabling measurements of the resistivity along the inter-chain ($a^*$) and interlayer ($c$) directions (see Fig.~\ref{fig:FIB_Crys}a). Remarkably, the anomaly is only pronounced in the transport channel perpendicular to the chains setting in at $T^*=20$\,K, but is quite subtle in all other directions. Finally, there is no change in the dHvA frequency observed across $T^*$, suggesting that the Fermi surface experiences no reconstruction at this temperature. We discuss possible explanations for the observed anomaly and its relationship to superconductivity, and in particular whether $T^*$ marks the onset of charge density wave (CDW) fluctuations or a thermodynamic phase transition corresponding to an incommensurate-to-commensurate CDW lattice lock-in.

\section{Experimental and Calculational Methods}
\label{exp}

Single crystals of TPT were grown using a self-flux with excess tellurium at temperatures as high as $1000$\,$^\circ$C, similar to the method reported by Jiao \emph{et al}.~\cite{jiao_superconductivity_2014}. High-resolution single-crystal X-ray diffraction was performed at the Advanced Light Source (ALS) at Lawrence Berkeley National Laboratory that confirmed the quality of the crystals and their orientation. Hence, contributions from impurity phases to the magnetic quantum oscillation data presented in the following can be ruled out.

\begin{figure}[tb]
\centering
\includegraphics[width=1.0\columnwidth]{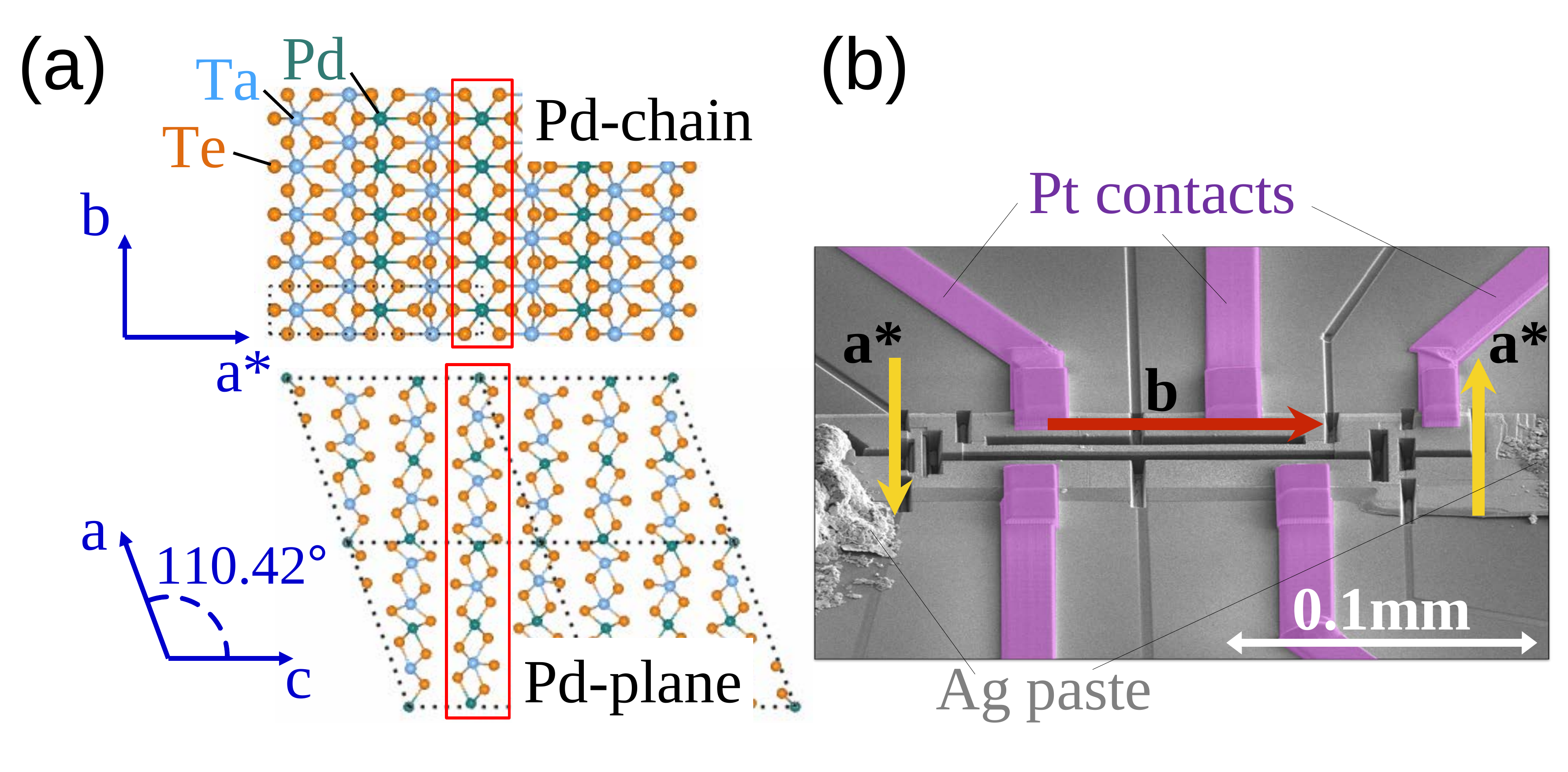}
\caption{{\bf (a)} Crystal structure of TPT with the conventional unit cell (dotted) in the representation C2/m. {\bf (b)} Scanning electron microscope image of a transport device prepared by FIB etching. Red and yellow arrows highlight the $b$- and $a^*$-directions.}
\label{fig:FIB_Crys} 
\end{figure}

Electrical transport devices were prepared by FIB etching and {\it in situ} deposition of platinum or gold contacts onto thin flakes of TPT, cleaved from the as-grown needle-like crystals. Sample sizes were a few micrometers in thickness and a couple of hundred micrometers in length. Low-ohmic contacts with contact resistances on the order of $1\Omega$ were achieved. A device that has been microstructured to measure $b$-axis and $a^*$-axis conductivity is shown in Fig.~\ref{fig:FIB_Crys}b.

Magnetic torque and magnetotransport experiments were performed in fields up to $45$\,T at the National High-Magnetic Field Laboratory (NHMFL) in Tallahassee, Florida. Both piezoresistive microcantilevers from SEIKO and capacitive cantilevers built by NHMFL were used as torque magnetometers. Further torque and magnetotransport measurements were performed in a Cryogenic Limited $16$\,T cryogen free magnet system.

Fully-relativistic first-principles calculations were performed using Density Functional Theory (DFT) under the local density approximation (LDA)~\cite{perdew1981} and implemented using the Quantum Espresso software~\cite{giannozzi2009}. Starting from the experimental crystal lattice parameters, all internal atomic positions were allowed to relax until interatomic forces were less than $5$\,meV$/$\AA. High-resolution Fermi surface data were obtained by non-self-consistent calculations on a $50\times50\times30$ $k$-point grid in the first Brillouin zone and were plotted using XCrySDen~\cite{Kokalj1999}. The dHvA simulation curves were calculated using the Supercell K-space Extremal Area Finder (SKEAF) code~\cite{rourke2012}.

\section{Magnetotransport and Magnetic Susceptibility}

\begin{figure}[tb]
\centering
\includegraphics[width=1.0\columnwidth]{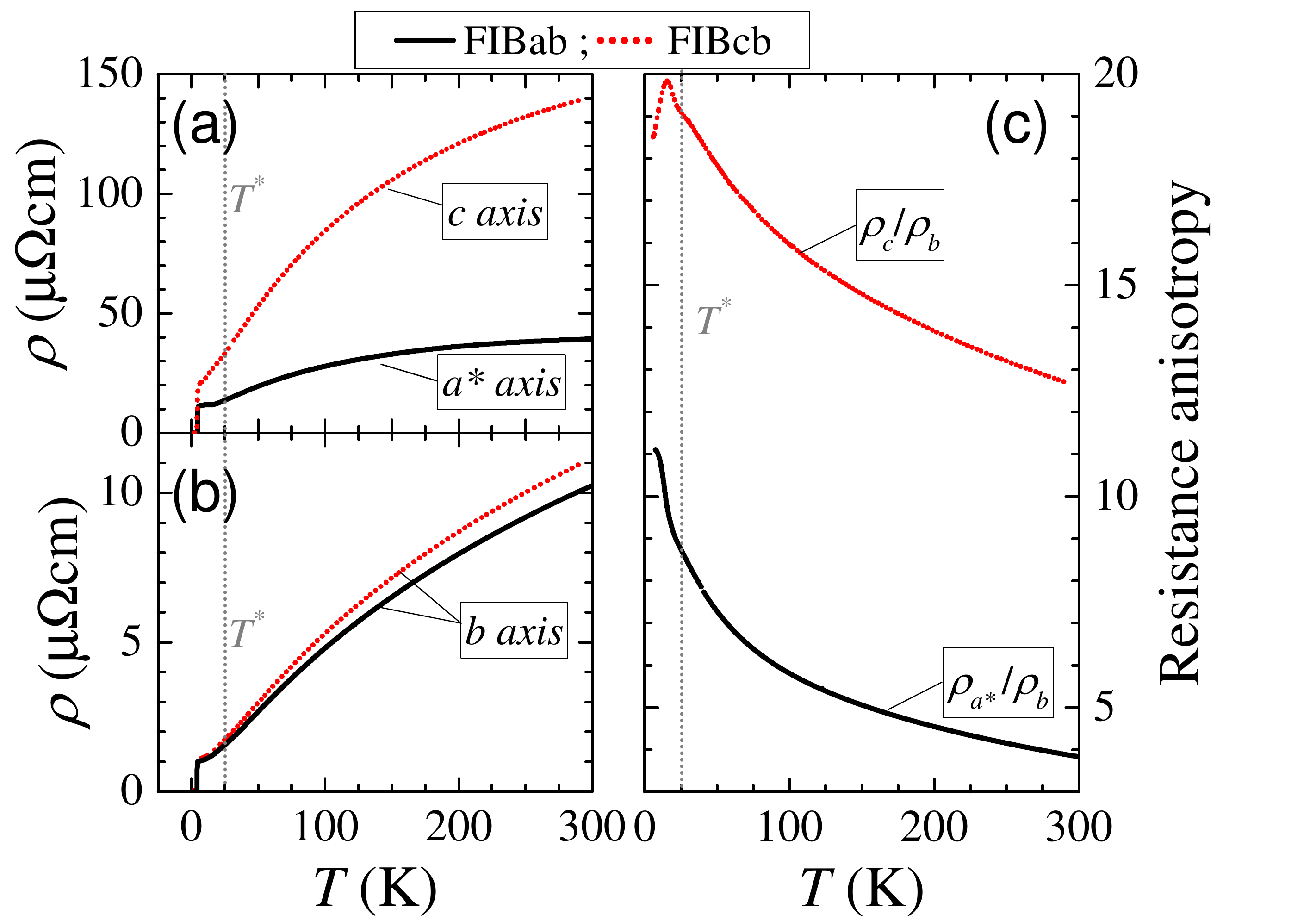}
\caption{{\bf(a)} and {\bf(b)} Resistivity along the $a^*$ and $b$ axes for device FIBab and along the $b$ and $c$ axes for device FIBcb plotted against temperature; the gray dotted line highlights the anomaly at $T^*$. {\bf(c)} Resistivity anisotropy $\rho_i / \rho_b$ for each device, with $i=a^*,c$.}
\label{fig:Rho_abc}
\end{figure}

Figures~\ref{fig:Rho_abc}a and \ref{fig:Rho_abc}b show the resistivity, measured with an applied current along the three crystallographic directions, as a direct measure of the transport anisotropy. We microstructured two devices: one made to measure the $a^*$ and $b$ directions simultaneously, denoted FIBab (shown in Fig.~\ref{fig:FIB_Crys}b), and the other measuring the $b$ and $c$ crystallographic directions simultaneously, denoted FIBcb. Note that FIBab and FIBcb, which are cut from crystals belonging to the same batch, have highly reproducible $b$-axis resistivity $\rho_b$, indicating homogeneity within the batch. The resistivity anisotropy grows monotonically as the temperature is lowered: at room temperature the resistivity ratios $(\rho_{a^*}:\rho_b:\rho_c) \sim 4:1:13$, while at low temperature (just above $T_c=4.6$\,K),  $(\rho_{a^*}:\rho_b:\rho_c) \sim 10:1:20$, as shown in Fig.~\ref{fig:Rho_abc}c. Due to possible mixing between different conduction channels, these ratios are lower bounds of the intrinsic resistivity anisotropies. Evidently, the system becomes much more conductive along the $b$-axis than along any other axis, suggesting that scattering becomes significantly suppressed along the chains. We find that both the temperature dependence of both $\rho_c/\rho_b$ and of $\rho_{a^*}/\rho_b$ show shallow kinks at $T^*=20$\,K followed by peaks above $T_c$.

\begin{center}
\begin{figure*}[tb]
\centering
        \includegraphics[width=\textwidth]{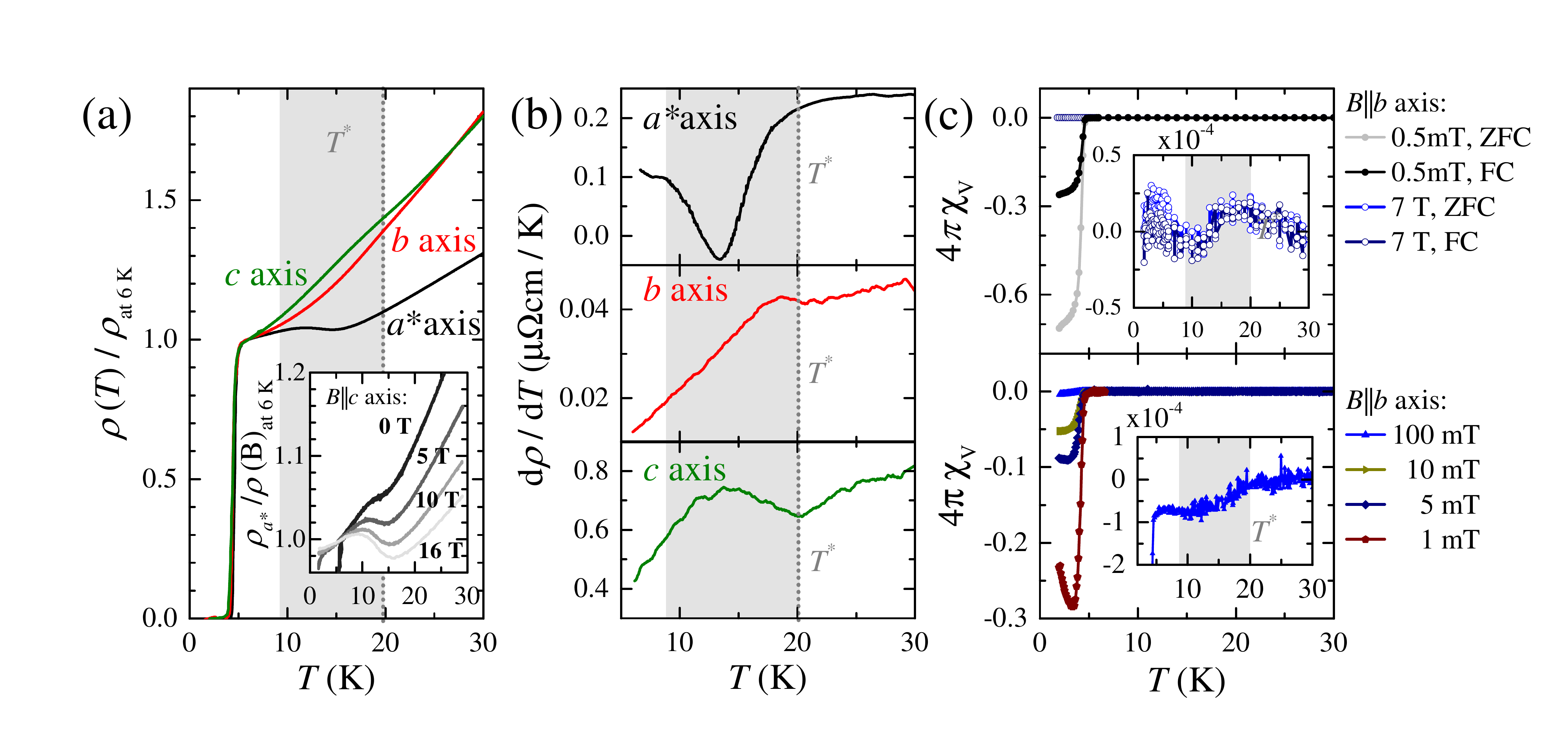}
    \caption{{\bf(a)} Resistivity (normalized by its value at $T=6$\,K) along all three crystal axes ($a^*$, $b$ and $c$) plotted against temperature; the gray dotted vertical line marks $T^*$ (see text). Inset: temperature dependence of the $a^*$-resistivity in a constant magnetic field up to $16$\,T applied parallel to $c$. {\bf(b)} Derivatives of the resistivity, $d\rho_i / dT$, for each transport direction, with $i=a^*,b,c$. {\bf(c)} Temperature dependence of the magnetic susceptibility $\chi_v$ for a small number of TPT co-aligned single crystals (upper panel) and a separated single crystal, oriented with the $b$ axis parallel to the applied magnetic field; insets show close-ups of the background subtracted high field data. Note: $T^*$ marks the temperature below which both resistivity and magnetization show a clear feature indicative of a thermodynamic phase transition;  The anomalous temperature region is highlighted by the grey shaded area.}
		\label{fig:deriv_mag}
\end{figure*}
\end{center}

As can be seen in Fig.~\ref{fig:deriv_mag}a, which shows the resistivity normalized to the $6$\,K value for all three channels, this peak is related to a resistivity anomaly which is most pronounced in $\rho_{a^*}$. In contrast, the data for $\rho_b$ and $\rho_c$ show very weak features in this region. On taking the derivative $d\rho_i/dT$ ($i\in (a^*,b,c)$), shown in Fig.~\ref{fig:deriv_mag}, we see that all crystallographic directions exhibit an anomaly within a few Kelvin of $T^*$. The grey shaded area in Fig.~\ref{fig:deriv_mag} highlights the temperature range, $T=[8-20]$\,K within which we observe deviations from normal metallic behavior. For the $a^*$-direction we find a clear minimum at $14$\,K but a broad onset starting slightly above $T^*$. Along the $b$-direction we find a weak change in slope accompanied by a peak slightly below $T^*$. Similarly, a peak develops for $d\rho_c/dT$ below $T^*$ with its maximum at $14$\,K. Given that the conductivity along $b$ is over ten times greater than that along $a^*$ at these temperatures (see Fig.~\ref{fig:Rho_abc}c) it seems unlikely that this feature is exclusively due to mixing of $\rho_b$ with $\rho_{a^*}$; rather, it is intrinsic, albeit weak enough to have been overlooked in previous measurements. We find that the $T^*$ anomaly is independent of the substrate (either sapphire or silicon), sample thickness (ranging from $15$\,$\mu$m to 1\,$\mu$m) and attachment method (either glue or van der Waals attraction). We can therefore rule out any substrate strain effects as a cause of the observed features.

The derivatives $d\rho_b/dT$ and  $d\rho_c/dT$ are strongly linear at $T<T^*$, consistent with the onset of $\rho\propto T^2$ Fermi liquid-like behavior, which may be a precursor to the onset of superconductivity at $T_c\sim4$\,K. Interestingly, the temperature at which the anomaly occurs is weakly sensitive to the application of magnetic field (inset Fig.~\ref{fig:deriv_mag}a), suggesting that it is strongly pinned electronically, or that it is related to the softening of a phonon mode. However, the magnetoresistance is much larger below the anomaly than above, suggesting the carriers are significantly more mobile at $T<T^*$. A similar response is observed in the unconventional superconductor BaFe$_2$As$_2$, where a strong scattering channel is gapped out at the N\'eel transition, strongly enhancing the mobility of the itinerant carriers~\cite{rotter2008}.

\begin{center}
\begin{figure*}[tb]
\centering
\includegraphics[width=1.0\textwidth]{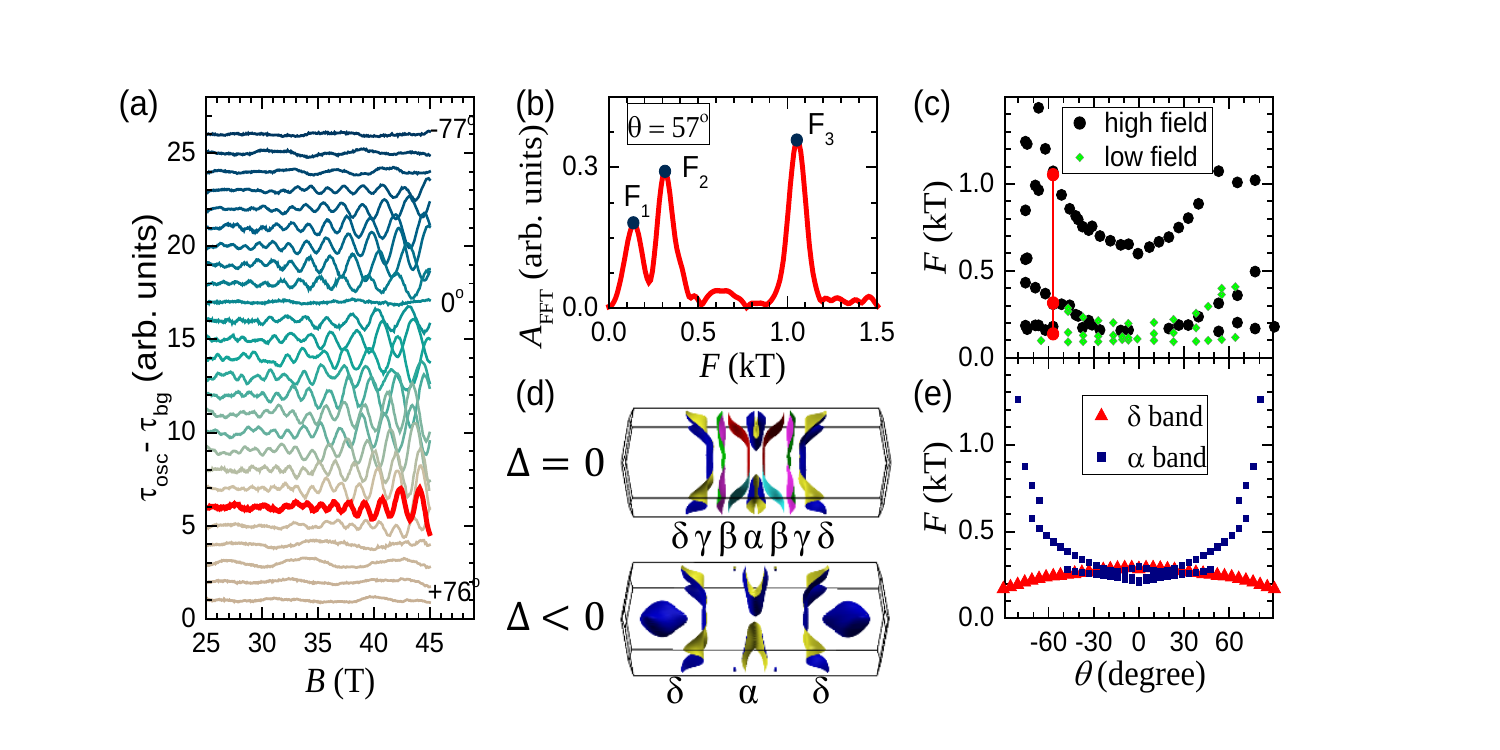}
\caption{{\bf(a)} Background-subtracted dHvA oscillation data for various field orientations recorded at $T=1.5$\,K in the $45$\,T hybrid magnet. $\theta=0^\circ$ corresponds to the applied field along the $c$-axis. {\bf(b)}, {\bf(c)} Frequency values obtained from fast Fourier transform (FFT) of the data in {\bf (a)}. The field ranges from $25-45\,$T (black spheres) and $5.5-16\,$T(green diamonds). {\bf(d)} Fermi surfaces for unshifted and shifted chemical potential with $\Delta=0$ and $-27$\,meV, respectively. {\bf(e)} DFT-calculated dHvA oscillation frequencies originating from the $\alpha$ and $\delta$ bands of the Fermi surfaces shown in {\bf(d)}.}
\label{fig:dHvA_45T} 
\end{figure*}
\end{center}

We studied the magnetic volume susceptibility $\chi_V$ of TPT around $T^*$ for a few dozen co-aligned single crystals with an average width of about $100\,\mu$m, thickness of $20\,\mu$m and length of $3$\,mm, pointing with their $b$ directions parallel to the field. In Fig.~\ref{fig:deriv_mag}c we show data measured at various magnetic fields, exhibiting strong diamagnetism setting in at the superconducting critical temperature $T_c=4.6$\,K, with a transition width of about $1$\,K. The inset of Fig.~\ref{fig:deriv_mag}c shows a zoom into the data recorded at a fields of $5$ and $7$\,T. Below $20$\,K there is a broad anomaly, manifested as a reduction in the magnetic susceptibility, followed by a subtle paramagnetic increase, before the superconducting transition sets in. Such a paramagnetic enhancement is a prominent feature known, for example from cuprate superconductors due to the onset of fluctuations in the vicinity of $T_c$ related to low dimensionality or granularity\cite{Wohlleben1991}. We also observed a similar decrease of $\chi_V$ for a separated single crystal (see insert in the lower panel of Fig.~\ref{fig:deriv_mag}c).

In the simplest model, the strong reduction above the superconducting transition can be attributed to a decrease in the electronic density of states or, equivalently, a decrease in the electronic effective mass (note that this is consistent with the enhanced mobility we observe). The consistency between transport and susceptibility data is strong evidence that the $T^*$ anomaly is a true thermodynamic transition, indicative of a substantial change in the electronic properties of the system.

\section{Magnetic Quantum oscillations and DFT band structure}

In order to better understand the low-temperature ground state, we study the Fermi surface by measuring magnetic quantum oscillations in high-purity crystals. Background-subtracted de Haas-van Alphen (dHvA) oscillations, recorded in a DC magnetic field at $T=1.5\,$K, are shown in Fig.~\ref{fig:dHvA_45T}a. Multiple frequencies superimposed on a quadratic-in-field magnetic background torque, $\tau_{bg}$ are observed. Fast Fourier transforms (FFTs) of the data for each angle $\theta$ ($\theta=0$ defined to have the applied magnetic field, $\mathbf{B}$, parallel to the $c$-axis) as shown in Fig.~\ref{fig:dHvA_45T}b for $\theta=57^\circ$, provide access to the angle dependence of the various frequencies. In Fig.~\ref{fig:dHvA_45T}c we show the angle dependence of the major frequency peaks for two separate experiments: in low fields, $B<16\,$T (green diamonds), and in fields between $25-45$\,T (black circles). In the high-field data at least three distinct frequencies are discernible, labeled $F_1$, $F_2$, and $F_3$. However, only $F_1$ and $F_2$ were discernible in fields below $16$\,T. We therefore screened three different growth batches and only observed contributions from $F_1$ and $F_2$. 

As the field orientation is tilted away from $\mathbf{B}||c$, the smallest frequency $F_1$ stays relatively unaffected, indicative of a three-dimensional (3D) Fermi surface. At $30$\,T the oscillation period for a frequency of about $100$\,T can be estimated to $\approx 9$\,T precision. Therefore, for a field window of $25-45$\,T at most four maxima can be observed, reducing the reliability of FFT analysis for such low frequencies. By contrast, $F_2$ and $F_3$ trace a trajectory that tracks the perpendicular component of field $1/B\cos{\theta}$, indicative of a quasi-two-dimensional (q2D) Fermi surface pocket. $F_3$ was only clearly visible in the high magnetic field data; it appeared close to the background noise level for the lowest temperatures in the experiments below $16$\,T. We reproduced the oscillation spectra for various samples by application of two distinct measurement techniques: capacitive and piezoresistive torque magnetometry. The clear presence of q2D and 3D Fermi surface components can explain the relatively weak anisotropic character of TPT's transport behavior compared to other q1D materials.

TPT's Fermi surface predicted by DFT calculations is shown in Fig.~\ref{fig:dHvA_45T}d. It consists of four distinct bands: a cylindrical $\alpha$ band centered around the $X$ point, and three q1D bands labeled $\beta$, $\gamma$, and $\delta$, respectively. The $\delta$ band shows a strong effect on the position of the chemical potential $\Delta$. By only minor changes in $\Delta$, equivalent to doping holes, this Fermi surface sheet acquires a more complex appearance exhibiting a three-dimensional extension along the $k_y$ direction in the Brillouin zone.

In Fig.~\ref{fig:dHvA_45T}e we show for comparison the results of DFT calculations for the Fermi surface at zero temperature. With a shift of the chemical potential by $\Delta\approx 27$\,meV, our calculations broadly agree with low-frequency oscillations originating from the cylindrical $\alpha$ and the 3D-part of the $\delta$ band. However, the higher-frequency oscillation $F_3$ cannot be simultaneously recovered, suggesting that the predicted band structure does not adequately capture the low-temperature ground state. This discrepancy could  be evidence for band reconstruction due to density-wave-broken translational symmetry. In addition, it is also possible that $F_3$ is a magnetic breakdown frequency that only appears at high magnetic fields and this is also consistent with a folded Fermi surface at low temperatures.

\onecolumngrid
\begin{center}
\begin{figure}[tb]
\centering
\includegraphics[width=1.0\textwidth]{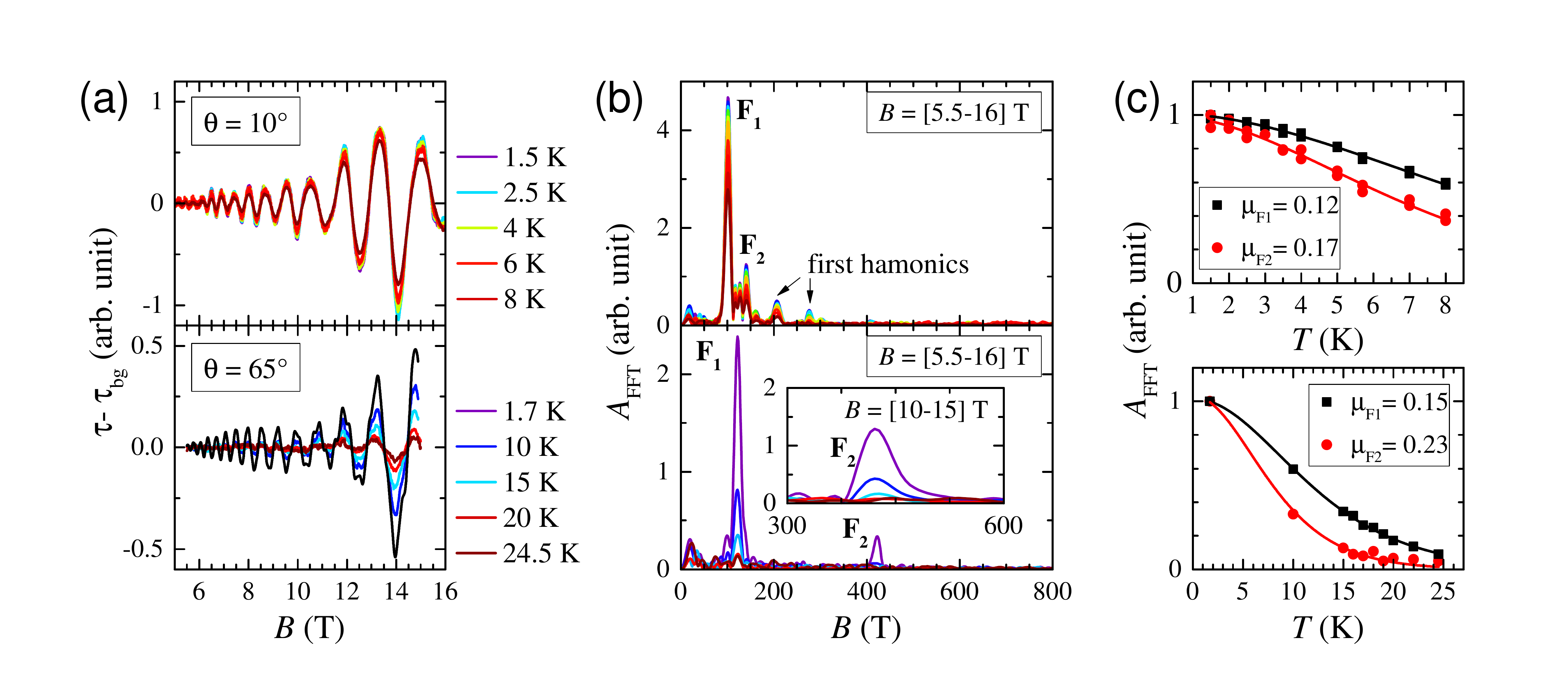}
\caption{{\bf (a)} Background-subtracted de Haas van Alphen oscillations measured with a piezoresistive torque magnetometer in a DC $16$\,T magnet for various temperatures, with $\theta=10^\circ$ in the upper panel and $\theta=65^\circ$ in the lower. {\bf (b)} FFT of the data shown in (a) over a field window of $[5.5-15]$\,T; The inset shows FFT data for $B=[10-15T]$\,T. {\bf (c)} Effective mass plots. Solid lines are fits according to Lifshitz-Kosevich theory (see main text).}
\label{fig:dHvA_16T} 
\end{figure}
\end{center}
\twocolumngrid
To investigate whether the Fermi surface is affected by the $T^*$ anomaly, we track the dHvA signal to higher temperatures (Fig.~\ref{fig:dHvA_16T} lower panels). We are able to resolve clear quantum oscillations in low fields below $16$\,T. The dHvA oscillation data are presented in Fig.~\ref{fig:dHvA_16T}a. Though the higher frequency is quickly suppressed with temperature, the 3D pocket at $\sim100\,$T can be tracked to $25\,$K (Fig.~\ref{fig:dHvA_16T}b). We observe no significant changes for this frequency in the FFT spectrum, suggesting that little or no band folding occurs at $T^*=20$\,K. Nevertheless it is possible that this pocket exists at a point in the Brillouin zone that is not reconstructed, this would be difficult to reconcile with our DFT calculations~\cite{shoenberg_1984}. One scenario, which we discuss below, is that $T^*$ marks an incommensurate-commensurate charge density wave lock-in transition, which requires no additional translational symmetry breaking.

The effective masses of each of the Fermi surface sheets are extracted from fits of the temperature-dependent oscillation amplitude at a constant field orientation to the Lifshitz-Kosevich temperature damping factor $R_T=K \mu T/B\sinh{(K \mu T/B)}$, with $K=-14.694$\,T$/$K~\cite{shoenberg_1984}. We find $\mu_{F_1}\approx (0.14\pm0.02) m_e$ and $\mu_{F_2}\approx (0.2\pm0.03) m_e$, with $m_e$ the bare electron mass. The fits are shown in Fig.~\ref{fig:dHvA_16T}c. The measured and calculated masses are quite similar, suggesting that correlation effects only weakly renormalize the electronic system. The difference between the two values supports the idea that the two Fermi surface sheets are located on different bands.

\section{Discussion}

At present, we cannot conclusively identify the microscopic nature of the thermodynamic anomaly at $T^*=20$\,K. It has long been suspected that owing to its nested Fermi surface, TPT could exhibit a Peierls instability. While low temperature scanning tunneling microscopy measurements appear consistent with this~\cite{fan_STM_2015}, there have been precious few measurements showing evidence of a transition~\cite{jiao_superconductivity_2014,pan_nodal_2014,jiao_multiband_2015,chen_raman_2015,Li_2016} at $T>T_c$, leaving open the interpretation of $T^*$ detected in this study.

On the one hand, it seems unlikely to be purely a CDW transition along the chain direction; the indifference of the quantum oscillation suggests there is no Fermi surface folding across $T^*$, there is no reported anomaly in heat capacity~\cite{pan_nodal_2014}, and in any case, the conductivity is enhanced along the chains, whereas one would naively expect it to be strongly suppressed due to the gapping out of the q1D Fermi surface. In the structurally similar rare earth tri-tellurides RTe$_3$ a CDW formation is accompanied by a notably greater conductivity reduction in their inter-chain direction (but this arises from a peculiarity in the Fermi surface geometry that is irrelevant in TPT~\cite{Monceau12,SinchenkoEA15}.) As a test we carried out conductivity calculations at zero temperature, similar to those employed for the RTe$_3$~\cite{SinchenkoEA15}, assuming a hypothetical band folding due to a CDW (as reported from STM~\cite{fan_STM_2015}). We however find that this single particle model fails to reproduce the observed enhanced effect on the transport channel perpendicular to the chains.

On the other hand, the $T^*$ anomaly may still be related to a Peierls-like Fermi instability, for example it could mark the onset of CDW fluctuations or an incommensurate (ICDW) to commensurate (CCDW) transition. In the former case, such an onset would give a resistivity bump to the $a^*$ direction just as we observe, arising from an enhancement in interchain interactions that lead to repulsion as each chain attempts to order. This is not unprecedented and indeed a similar mechanism as has been observed in the q2D system P$_4$W$_{10}$O$_{38}$~\cite{SchlenkerEA96,NaitoTanaka82} and predicted for NbSe$_2$~\cite{FFJvW_NatComms15,FFJvW_PRB15}.Despite the fluctuating CDW order being along the $\mathbf{b}$-direction chains it would not necessarily give a strong enhancement of the resistivity in this direction. The three-dimensional character of the fluctuations could plausibly have a stronger effect on the inter-chain direction as instantaneous regions of the CDW attempt to order the charge on the neighboring chains. However, this mechanism doesn't answer the apparent disagreement between the DFT calculations and our observed QOs.

ICDW to CCDW transitions can be driven by an electron-phonon interaction as the ICDW locks into the lattice~\cite{McMillan76,FFJvW_PRL15}, and is known to occur in other compounds including TTF-TCNQ, $2H$-TaSe$_2$ and $2H$-TaS$_2$~\cite{KagoshimaEA76,NaitoTanaka82,Monceau12,FengEA15}. This could account for all our data and was consistent with recent nuclear quadrupole resonance measurements which show a line broadening at $T\approx T^*$, but importantly no new lines (as would be expected for a singular CDW transition). However, this does raise the question of the location of the ICDW transition temperature, which we do not detect. This may be because $T_{ICDW}$ is higher than the range measured, or sufficiently subtle that we have not been able to detect it. In this case, we suggest that high-resolution x-ray measurements may be able to detect the ICDW, as well as resolve the structural nature of the $T^*$ anomaly.

\section{Conclusions}

Finally, we point out that there are other possibilities that could be the low temperature ground state (such as a spin-density wave, nematic transition or some other broken symmetry). Nevertheless, we can say with certainty that there is some kind of thermodynamic transition at $T^*$. Intriguingly, this both cuts off the growth of the transport anisotropy that was growing precipitously until $T^*$ (Fig.~\ref{fig:Rho_abc}c), as well as leading to a monotonic $T^2$ dependence before going superconducting (Fig.~\ref{fig:deriv_mag}b). It may therefore we inferred that while the thermodynamic nature of $T^*$ may not be directly driving $T_c$ in TPT, it may `set the scene' for conventional superconducting behavior to come about.

\section{Acknowledgements}
We would like to acknowledge the extremely helpful and efficient collaboration of S.~J.~Teat and K.~J.~Gagnon for characterizing and orienting the single crystals at beam line $11.3.1$ of the Advanced Light Source (ALS) in Berkeley in preparation of the individual experiments. We are grateful for the help of A. Frano in terms of performing measurements of the magnetic susceptibility in a Quantum design SQUID magnetometer in the laboratory of R. J. Birgeneau. We thank K.~R.~Shirer and M.~Baenitz for stimulating discussions. The ALS is supported by the Director, Office of Science, Office of Basic Energy Sciences, of the U.~S.~Department of Energy under Contract No.~DE-AC02-05CH11231. T.~H. was supported by the Quantum Materials FWP, U.S. Department of Energy, Office of Basic Energy Sciences, Materials Sciences and Engineering Division, under Contract No.~DE-AC02-05CH11231. F.~F.~acknowledges support from a Lindemann Trust Fellowship of the English Speaking Union. L.~B.~is supported by DOE-BES through award DE-SC0002613.
\clearpage

\clearpage
%%%%%%%%%%%%%%%%%%%%%%%%%%%%%%%%%%%%%%%%%%%%%
\bibliographystyle{unsrt}
\bibliography{TPT_Helm_bib}

\begin{thebibliography}{10}

\bibitem{Hinkov_2008}
V.~Hinkov, D.~Haug, B.~Fauqu{\'e}, P.~Bourges, Y.~Sidis, A.~Ivanov,
  C.~Bernhard, C.~T. Lin, and B.~Keimer.
\newblock Electronic {Liquid} {Crystal} {State} in the {High}-{Temperature}
  {Superconductor} {Y}{Ba}$_2${Cu}$_3${O}$_{6.45}$.
\newblock {\em Science}, 319(5863):597--600, 2008.

\bibitem{Fernandes_2014}
R.~M. Fernandes, A.~V. Chubukov, and J.~Schmalian.
\newblock What drives nematic order in iron-based superconductors?
\newblock {\em Nat. Phys.}, 10:97--104, 2014.

\bibitem{Wu2011}
T.~Wu, H.~Mayaffre, S.~Kramer, M.~Horvartic, C.~Berthier, W.~N. Hardy,
  R.~Liang, D.~A. Bonn, and M.-H. Julien.
\newblock Magnetic-field-induced charge-stripe order in the high-temperature
  superconductor {YB}a$_2${C}u$_3${O}$_y$.
\newblock {\em Nature}, 477:191--194, 2011.

\bibitem{Christensen2016}
Morten~H. Christensen, Henrik Jacobsen, Thomas~A. Maier, and Brian~M. Andersen.
\newblock Magnetic fluctuations in pair-density-wave superconductors.
\newblock {\em Phys. Rev. Lett.}, 116:167001, Apr 2016.

\bibitem{badoux2016}
S.~Badoux, W.~Tabis, F.~Laliberté, G.~Grissonnanche, B.~Vignolle,
  D.~Vignolles, J.~Béard, D.~A. Bonn, W.~N. Hardy, R.~Liang,
  N.~Doiron-Leyraud, Louis Taillefer, and Cyril Proust.
\newblock Change of carrier density at the pseudogap critical point of a
  cuprate superconductor.
\newblock {\em Nature}, 531(7593):210--214, 2016.

\bibitem{Peierls1930}
R.~Peierls.
\newblock Zur {Theorie} der elektrischen und thermischen {Leitfähigkeit} von
  {Metallen}.
\newblock {\em Ann. Phys.}, 396(2):121--148, 1930.

\bibitem{Pouget2016}
J.-P. Pouget.
\newblock The {Peierls} instability and charge density wave in one-dimensional
  electronic conductors.
\newblock {\em C. R. Physique}, 17(3–4):332 -- 356, 2016.

\bibitem{pan_nodal_2014}
J.~Pan, W.~H. Jiao, X.~C. Hong, Z.~Zhang, L.~P. He, P.~L. Cai, J.~Zhang, G.~H.
  Cao, and S.~Y. Li.
\newblock Nodal superconductivity and superconducting dome in the layered
  superconductor {Ta}$_4${Pd}$_3${Te}$_{16}$.
\newblock {\em Phys. Rev. B}, 92:180505, Nov 2015.

\bibitem{jiao_superconductivity_2014}
W.-H. Jiao, Z.-T. Tang, Y.-L. Sun, Y.~Liu, Q.~Tao, C.-M. Feng, Y.-W. Zeng,
  Z.-A. Xu, and G.-H. Cao.
\newblock Superconductivity in a layered {Ta}$_4${Pd}$_3${Te}$_{16}$ with
  pdte$_2$ chains.
\newblock {\em J. Am. Chem. Soc.}, 136(4):1284--1287, January 2014.

\bibitem{du_anisotropic_2015}
Z.~Du, D.~Fang, Z.~Wang, Y.~Li, G.~Du, H.~Yang, X.~Zhu, and H.-H. Wen.
\newblock Anisotropic {Superconducting} {Gap} and {Elongated} {Vortices} with
  {Caroli}-{De} {Gennes}-{Matricon} {States} in the {New} {Superconductor}
  {Ta}$_4${Pd}$_3${Te}$_{16}$.
\newblock {\em Sci. Rep.}, 5, 2015.

\bibitem{mar_synthesis_1991}
A.~Mar and J.~A. Ibers.
\newblock Synthesis, crystal structure and electrical conductivity of a new
  layered ternary telluride {Ta}$_4${Pd}$_3${Te}$_{16}$.
\newblock {\em J. Chem. Soc., Dalton Trans.}, S(S):639--641, 1991.

\bibitem{Dressel2007}
M.~Dressel.
\newblock Ordering phenomena in quasi-one-dimensional organic conductors.
\newblock {\em Naturwissenschaften}, 94(7):527--541, 2007.

\bibitem{Wakeham2011}
N.~Wakeham, A.~F. Bangura, X.~Xu, J.~F. Mercure, M.~Greenblatt, and N.~E.
  Hussey.
\newblock Gross violation of the wiedemann-franz law in a quasi-one-dimensional
  conductor.
\newblock {\em Nat. Comm.}, 2:396, July 2011.

\bibitem{singh_multiband_2014}
D.~J. Singh.
\newblock Multiband superconductivity of {Ta}$_4${Pd}$_3${Te}$_16$ from {Te} p
  states.
\newblock {\em Phys. Rev. B}, 90(14):144501, October 2014.

\bibitem{perdew1981}
J.~P. Perdew and Alex Zunger.
\newblock Self-interaction correction to density-functional approximations for
  many-electron systems.
\newblock {\em Phys. Rev. B}, 23(10):5048--5079, 1981.

\bibitem{giannozzi2009}
P.~Giannozzi, S.~Baroni, N.~Bonini, M.~Calandra, R.~Car, C.~Cavazzoni,
  D.~Ceresoli, G.~L. Chiarotti, M.~Cococcioni, I.~Dabo, A.~D. Corso, S.~d.
  Gironcoli, S.~Fabris, G.~Fratesi, R.~Gebauer, U.~Gerstmann, C.~Gougoussis,
  A.~Kokalj, M.~Lazzeri, L.~Martin-Samos, N.~Marzari, F.~Mauri, R.~Mazzarello,
  S.~Paolini, A.~Pasquarello, L.~Paulatto, C.~Sbraccia, S.~Scandolo,
  G.~Sclauzero, A.~P. Seitsonen, A.~Smogunov, P.~Umari, and R.~M. Wentzcovitch.
\newblock {QUANTUM} {ESPRESSO}: a modular and open-source software project for
  quantum simulations of materials.
\newblock {\em J. Phys.: Condens. Matter}, 21(39):395502, 2009.

\bibitem{Kokalj1999}
Anton Kokalj.
\newblock {XCrySDen}—a new program for displaying crystalline structures and
  electron densities.
\newblock {\em J. Mol. Graph. Mod.}, 17(3–4):176--179, 1999.

\bibitem{rourke2012}
P.~M.~C. Rourke and S.~R. Julian.
\newblock Numerical extraction of de {Haas}–van {Alphen} frequencies from
  calculated band energies.
\newblock {\em Compput. Phys. Commun.}, 183(2):324--332, 2012.

\bibitem{rotter2008}
M.~Rotter, M.~Tegel, D.~Johrendt, I.~Schellenberg, W.~Hermes, and R.~Pöttgen.
\newblock Spin-density-wave anomaly at $140$ k in the ternary iron arsenide
  {BaFe}$_2${As}$_2$.
\newblock {\em Phys. Rev. B}, 78(2):020503, 2008.

\bibitem{Wohlleben1991}
D.~Wohlleben, M.~Esser, P.~Freche, E.~Zipper, and M.~Szopa.
\newblock Possibility of orbital magnetic phase transitions in mesoscopic
  metallic rings.
\newblock {\em Phys. Rev. Lett.}, 66:3191--3194, Jun 1991.

\bibitem{shoenberg_1984}
D.~Shoenberg.
\newblock {\em Magnetic Oscillations in Metals}.
\newblock Cambridge University Press, 1984.
\newblock Cambridge Books Online.

\bibitem{fan_STM_2015}
Q.~Fan, W.~H. Zhang, X.~Liu, Y.~J. Yan, M.~Q. Ren, M.~Xia, H.~Y. Chen, D:~F.
  Xu, Z.~R. Ye, W.~H. Jiao, G.~H. Cao, B.~P. Xie, T.~Zhang, and D.~L. Feng.
\newblock Scanning tunneling microscopy study of superconductivity, magnetic
  vortices, and possible charge-density wave in {Ta}$_4${Pd}$_3${Te}$_{16}$.
\newblock {\em Phys. Rev. B}, 91:104506, 2015.

\bibitem{jiao_multiband_2015}
W.-H. Jiao, Y.~Liu, Y.-K. Li, X.-F. Xu, J.-K. Bao, C.-M. Feng, S.~Y. Li, Z.-A.
  Xu, and G.-H. Cao.
\newblock Multiband superconductivity in {Ta}$_4${Pd}$_3${Te}$_{16}$ with
  anisotropic gap structure.
\newblock {\em J. Phys.: Condens. Matter}, 27(32):325701, 2015.

\bibitem{chen_raman_2015}
D.~Chen, P.~Richard, Z.-D. Song, W.-L. Zhang, S.-F. Wu, W.~H. Jiao, Z.~Fang,
  G.-H. Cao, and H.~Ding.
\newblock Raman scattering investigation of the quasi-one-dimensional
  superconductor {Ta}$_4${Pd}$_3${Te}$_{16}$.
\newblock {\em J.Phys.: Condensed Matter}, 27(49):495701, 2015.

\bibitem{Li_2016}
Z.~Li, W.~H. Jiao, G.~H. Cao, and G.~Zheng.
\newblock {Charge fluctuations and nodeless superconductivity in
  quasi-one-dimensional {Ta}$_4${Pd}$_3${Te}$_{16}$ revealed by
  $^{125}${Te}-{NMR} and $^{181}${Ta}-{NQR}}.
\newblock In {\em eprint arXiv:1610.01811v1}, 2016.

\bibitem{Monceau12}
P.~Monceau.
\newblock Electronic crystals: an experimental overview.
\newblock {\em Advances in Physics}, 61, 2012.

\bibitem{SinchenkoEA15}
A.~Sinchenko, P.~D. Grigoriev, P.~Lejay, O.~Leynaud, and P.~Monceau.
\newblock Anisotropy of conductivity in rare-earth tritellurides in the static
  and sliding states of the {CDW}.
\newblock {\em Physica B}, 460:21--25, 2015.

\bibitem{SchlenkerEA96}
C.~Schlenker, C.~Hess, C.~Le~Touze, and J.~Dumas.
\newblock Charge {Density} {Wave} {Properties} of {Quasi} {Low-Dimensional}
  {Transition} {Metal} {Oxide} {Bronzes}.
\newblock {\em J. de Physique I}, 6:2061--2078, 1996.

\bibitem{NaitoTanaka82}
M.~Naito and S.~Tanaka.
\newblock Electrical {Transport Properties} in $2${H}-{NbS}$_2$, -{NbSe}$_2$,
  -{TaS}$_2$ and -{TaSe}$_2$.
\newblock {\em J. Phys. Soc. Jap.}, 51(1):219--227, 1982.

\bibitem{FFJvW_NatComms15}
F.~Flicker and J.~van Wezel.
\newblock Charge {Order} from {Orbital} {Dependent} {Coupling} {Evidenced} by
  {NbSe}$_2$.
\newblock {\em Nat. Commun.}, 6:7034, 2015.

\bibitem{FFJvW_PRB15}
F.~Flicker and J.~van Wezel.
\newblock Charge ordering geometries in uniaxially strained {NbSe}$_{2}$.
\newblock {\em Phys. Rev. B}, 92:201103, 2015.

\bibitem{McMillan76}
W.~L. McMillan.
\newblock Theory of discommensurations and the commensurate-incommensurate
  charge-density-wave phase transition.
\newblock {\em Phys. Rev. B}, 14:1496--1502, 1976.

\bibitem{FFJvW_PRL15}
F.~Flicker and J.~van Wezel.
\newblock {One-Dimensional} {Quasicrystals} from {Incommensurate Charge Order}.
\newblock {\em Phys. Rev. Lett.}, 115:236401, 2015.

\bibitem{KagoshimaEA76}
S.~Kagoshima, T.~Ishiguro, and A.~Hiroyuki.
\newblock X-ray scattering studies of phonon anomalies and superstructures in
  {TTF-TCNQ}.
\newblock {\em J. Phys. Soc.Jap.}, 41(6), 1976.

\bibitem{FengEA15}
Y.~Feng, J.~van Wezel, J.~Wang, F.~Flicker, D.~M. Silevitch, P.~B. Littlewood,
  and T.~F. Rosenbaum.
\newblock Itinerant density wave instabilities at classical and quantum
  critical points.
\newblock {\em Nat. Phys.}, 11:865, 2015.

\end{thebibliography}
%%%%%%%%%%%%%%%%%%%%%%%%%%%%%%%%%%%%%%%%%%%%%

%\appendix

\end{document}